\input{aipcheck}

\documentclass[
    ,final            
  ]
  {aipproc}

\layoutstyle{8x11single}

\begin{document}

\title{Correlations between Lag, Duration, Peak Luminosity, \\ Hardness, and Asymmetry in Long GRB Pulses}


\author{Jon Hakkila}{
  address={Department of Physics and Astronomy, College of Charleston, Charleston, SC}
}

\author{Renata S. Cumbee}{
  address={Department of Physics and Astronomy, Francis Marion University, Florence, SC}
}


\begin{abstract}
Continued study of the BATSE catalog verifies previously-identified correlations between pulse lag and pulse duration and corresponding anti-correlations between both properties and pulse peak flux for a large sample of Long GRB pulses; the study also finds correlations between pulse peak lags, pulse asymmetry, and pulse hardness. These correlations apparently can be used to delineate Long GRBs from Short ones. Correlated pulse properties represent constraints that can be used to guide theoretical modeling, whereas bulk prompt emission properties appear to be constructed by combining and smearing out pulse characteristics in ways that potentially lose valuable information. 
\end{abstract}

\maketitle


\section{Introduction}

Important correlative characteristics have been attributed over the years to pulses in gamma-ray burst (GRB) prompt emission. Examples of these characteristics include (1) temporal asymmetry characterized by longer decay than rise rates, (2) hard-to-soft spectral evolution, and (3) broadening at lower energies (e.g. \cite{nor96,rrf00,nor02,ryd05}). Despite these suggestive correlations, GRB prompt emission studies have focused on the measurement of bulk properties, with pulses often regarded as nothing more than minor variations to the overall signal. 

Recently, Hakkila et al. \cite{hak08a, hak08b} demonstrated that bulk GRB characteristics are composite properties formed from characteristics belonging to distinct yet unresolved pulses. Each pulse in a small sample of GRBs with known redshift was found to have its own lag, and the lag associated with the bulk gamma-ray emission (using the cross correlation function \cite{ban97}) was shown to be a heterogeneous composite of the individual pulse lags favoring the high intensity, short duration pulses. Since bulk GRB lags are correlated with GRB peak luminosities \cite{nor00}, this implies that pulse properties must be highly correlated to produce measurable correlations in bursts composed of often many pulses. Pulses belonging to the small sample of BATSE GRBs with known redshifts used in this study have unique, individual pulse lags that correlate with the pulse durations and that anti-correlate with pulse peak luminosity.  In order to verify that this is not a statistical discrepancy unique to this data set, we have set out to expand our GRB pulse database by sequentially analyzing bursts in the Current BATSE Catalog (\url{http://www.batse.msfc.nasa.gov/batse/grb/catalog/current/}). The results allow us to determine how pervasive correlated pulse properties are while sampling the BATSE Catalog data in a relatively unbiased way.

\section{Procedure}

We use a semi-automated procedure and code to identify and fit pulses in multi-channel observations of GRB prompt emission using 64 ms data \cite{hak08a, hak08b}. The approach starts with a search for potential pulses in summed four channel data. Candidate time intervals potentially containing pulses are identified using the Bayesian Blocks methodology \cite{sca98}; a pulse model is optimized within each interval to see if it contains a statistically-significant pulse. The data from intervals containing statistically-insignificant model pulses are merged into the surrounding intervals, and the fitting process begins again. Iteration eventually produces an optimal set of model pulses for the summed four-channel data.

The four-parameter pulse model used \cite{nor05} assumes the following functional form for each pulse:
\begin{equation}
I(t) = A \lambda \exp^{[-\tau_1/(t - t_s) - (t - t_s)/\tau_2]},
\end{equation}
where $t$ is time since trigger, $A$ is the pulse amplitude, $t_s$ is the pulse start time, 
$\tau_1$ and $\tau_2$ are characteristics of the pulse rise and pulse decay, and 
$\lambda = \exp{[2 (\tau_1/\tau_2)^{1/2}]}$). A two-parameter background model is 
simultaneously fitted along with any pulses.
The resulting 4-channel pulse characteristics are used as starting points from
which individual energy channel pulse fits are obtained.
The aforementioned process is repeated until convergent solutions 
are obtained for pulses in each energy channel.

A variety of model-dependent pulse properties are extracted from the aforementioned pulse parameters in the fitting process; these include pulse durations, pulse peak fluxes (256 ms timescale), pulse peak lags, pulse asymmetries, and pulse spectral hardnesses. The {\em pulse duration} $w$ is obtained from the summed four-channel data and is defined as $w = [9 + 12\sqrt{\tau_1/\tau_2}]^{1/2}$; this is the interval between times when the pulse amplitude is $A {e^{-3}}$. The {\em pulse peak flux} $p_{256}$ is defined on the 256 ms timescale in terms of the summed four-channel data. {\em Pulse peak lags} $l$ are the differences between the pulse peak times in different energy channels (pulse peak times are given by $\tau_{\rm peak} = t_s + \sqrt{\tau_1 \tau_2}$). Pulse peak lags can be obtained for any pulse between two energy channels, although we define the standard pulse peak lag $l_{31}$ as that measured between energies of 100 to 300 keV (BATSE channel 3) and 25 to 50 keV (BATSE channel 1). The pulse asymmetry $\kappa$ is defined as $\kappa = w/(3 + 2\sqrt{\tau_1/\tau_2}).$ {\em Hardness ratios} $HR$ are constructed by dividing pulse fluences in two different energy channels; we use a hardness ratio $HR_{31}$ defined by $HR_{31}=S_3/S_1$, where $S_3$ is the channel 3 fluence and $S_1$ is the channel 1 fluence.

This analysis approach has been successfully applied to both BATSE and Swift bursts. However, difficulty in normalizing BAT data to BATSE data has thus far prevented us from merging measurements from the two datasets.

The procedure is fairly successful at pulse extraction, even though the pulse signal-to-noise ratio is often low in more than one of the BATSE energy channels (most often in channel 4). Pulse properties are cleanly extracted in a relatively unambiguous manner for GRBs containing non-overlapping or isolated pulses. Typically, these tend to be low luminosity, long duration bursts \cite{nor05, hak07}. However, the process is also successful at identifying and fitting many pulses in complex GRBs containing overlapping pulses. The approach is less successful at fitting low intensity pulses, pulses that strongly overlap, and very short pulses, which can be indistinguishable from Poisson noise. Ambiguous pulse identifications often result in poor $\chi^2$ goodness-of-fit measures, in fits that appear to merge pulses separable to the eye, and/or in pulses that have disparate properties or are not observed in contiguous energy channels. We exclude from our analysis  overlapping and low fluence pulses that are overtly ambiguous, but recognize existence of the latter by tentatively identifying them as low fluence events (LFEs).

We also classify the bursts that we study. Although the definition of the Short and Long class of GRBs has become more complex during the Swift era than it was, we use a classification scheme based on duration and spectral hardness originally obtained from BATSE bursts using machine learning algorithms \cite{hak07}. This scheme \citep{hak03} classifies GRBs as Short if they satisfy the inequality  (T90 $<1.954$) OR ($1.954 \le T90 < 4.672$ AND $HR_{321} > 3.01$), where $HR_{321} = S_3/(S_2 + S_1)$ \citep{muk98}, where $S_2$ is the channel 2 fluence. If GRBs do not satisfy the aforementioned inequality, then they are classified as Long.

When pulses have been fitted and the GRBs to which they belong have been classified, we are able to seek out possible correlations among the properties of pulses in Long and Short BATSE GRBs.

\section{Analysis}

Our current database consists of 307 pulses in 106 Long and 46 Short GRBs. We have started simultaneously at the ends of the Current BATSE Catalog and are working our way through the rest of the Catalog sequentially. The results obtained with this larger database are consistent with those obtained previously for smaller datasets \cite{hak08a, hak08b}: pulse properties, rather than bulk properties of the prompt emission, underly GRB measurements. 

Figure 1 demonstrates an example of pulse fits for BATSE Trigger 0214. This is a Long GRB with 
three fitted pulses. Pulse properties are extracted for this and many other BATSE burst, and compiled in a pulse database.  

\begin{figure}
  \includegraphics[height=.27\textheight]{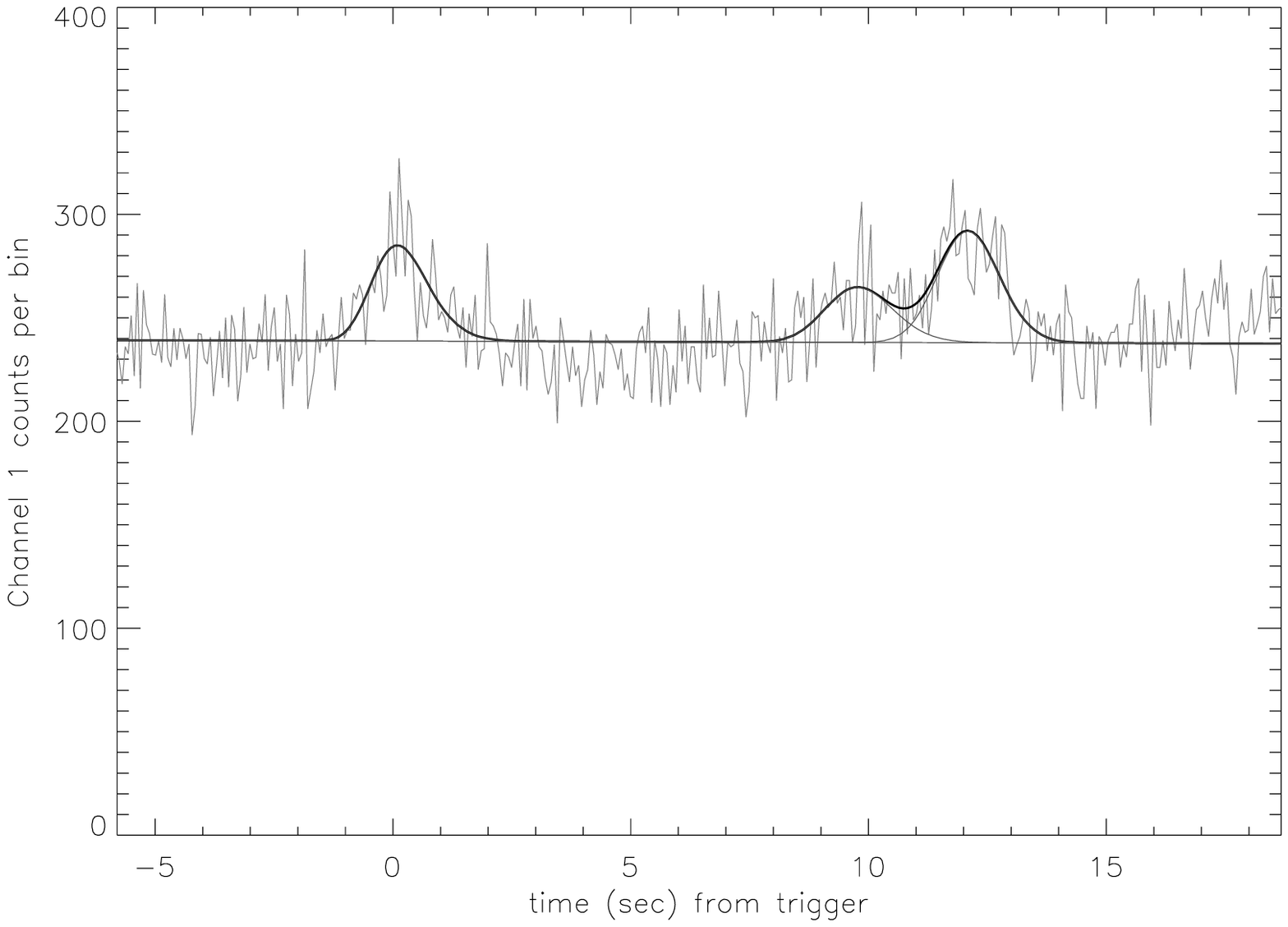}
  \includegraphics[height=.27\textheight]{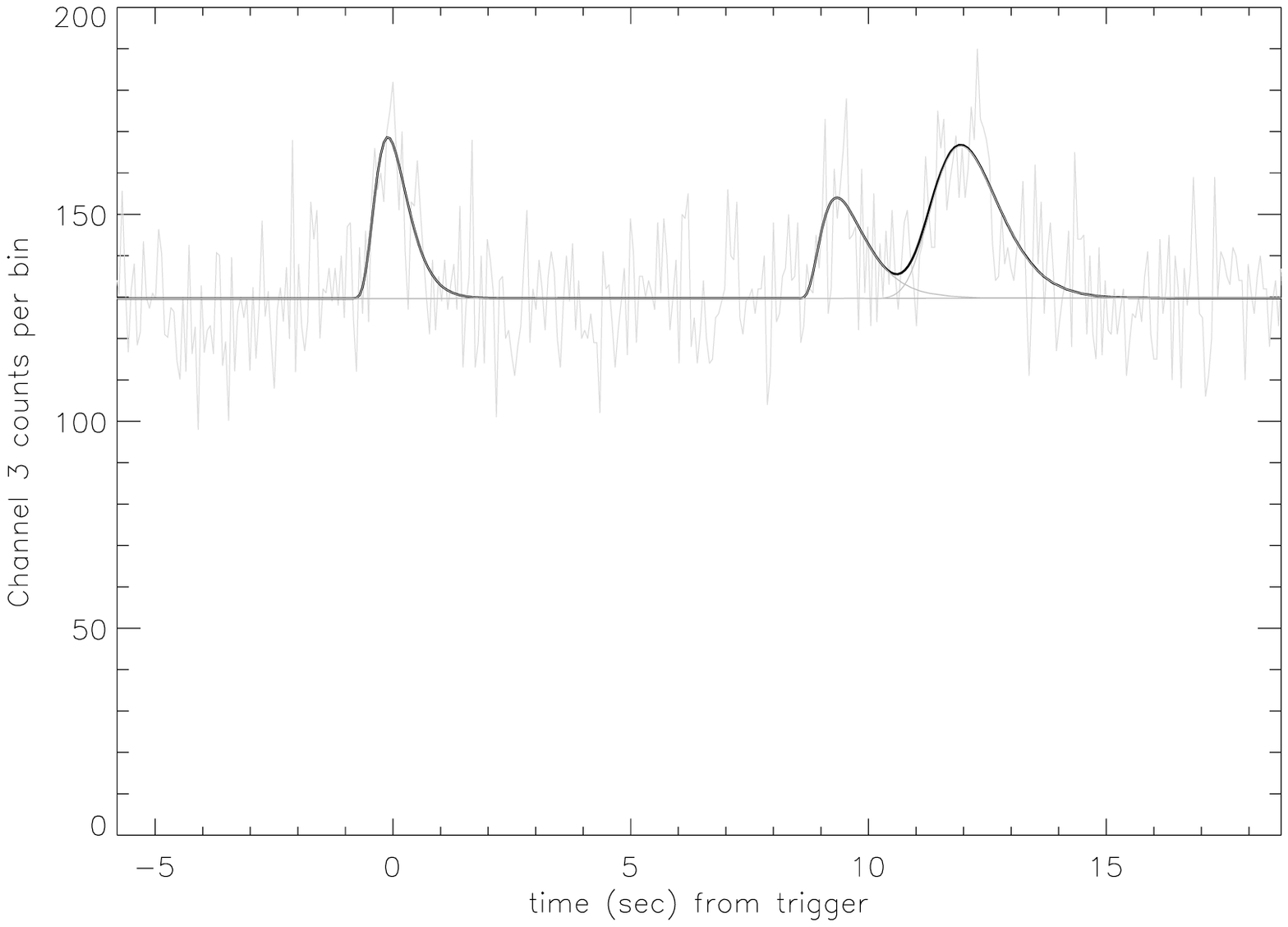}
  \caption{Example of pulse fits for a Long BATSE GRB (Trigger 0214): (a) channel 1, and (b) channel 3. The pulses have the following lags: pulse 1 lag = 0.679 s, pulse 2 lag = 0.368 s, and pulse 3 lag = 0.352 s. }
\end{figure}

\subsection{Correlations}

Pulse properties are compared and correlated for both the Long and Short GRB classes. The most rigorous correlation found is the one between pulse duration and pulse peak lag: the probability that these are uncorrelated in Long GRB pulses is $2.3 \times 10^{-19}$. Such a strong correlation suggests that the pulse duration and the pulse peak lag are two different representations of the same phenomenon; these also appear related to hard-to-soft pulse evolution. Both characteristics have been previously found to anti-correlate with pulse peak luminosity \cite{hak08a} and are thus luminosity indicators. We note that it is much easier to accurately measure pulse durations than pulse peak lags, since pulse durations are longer and since the smoothly-varying pulse fitting function allows the pulse duration to be accurately determined.

Although the pulse duration vs.\ pulse peak lag relation is generally very tight, there are pulses with specific pulse properties for which this relation may not correlate. Pulse peak lags of Short GRB pulses, for example, do not strongly correlate with pulse durations; this could be because it is difficult to measure pulse peak lags for Short GRB pulses using the 64 ms timescale. The lags of Short GRB pulses are so short that roughly half of them are negative; this is consistent with a distribution centered around zero seconds and is likely due primarily to measurement error. Additionally, a small number of Long GRB pulses are found to have short lags but long durations; these appear to be long pulses with sharply-defined pulse peaks. These pulses also appear to be inconsistent with the Pulse Scale Conjecture of Nemiroff \cite{nem00} which assumes that all pulses have temporal structures that are similar when properly scaled. Hakkila et al. \cite{hak08a} have previously suggested that these pulses might be external shock signatures, but it is possible that these represent inaccurately-measured prorpeties of overlapping pulses.

We note that many Long GRBs contain short duration pulses; these typically have pulse peak lags near zero seconds. This indicates that a short duration pulse does not necessarily identify a GRB belonging to the Short class. Likewise, there are many Long GRB pulses with pulse lags that are demonstrably negative. Figure 3 (BATSE trigger 1807) is one of these: the negative-lag pulse does not appear to be a single pulse with homogeneous properties, but instead has two peaks indicative of overlapping pulses. Given the relative rarity of negative-lag pulses, we suggest that many of these are merely unresolved overlapping pulses, although some could be pulses of short duration whose lags cannot be accurately measured.

\begin{figure}
  \includegraphics[height=.27\textheight]{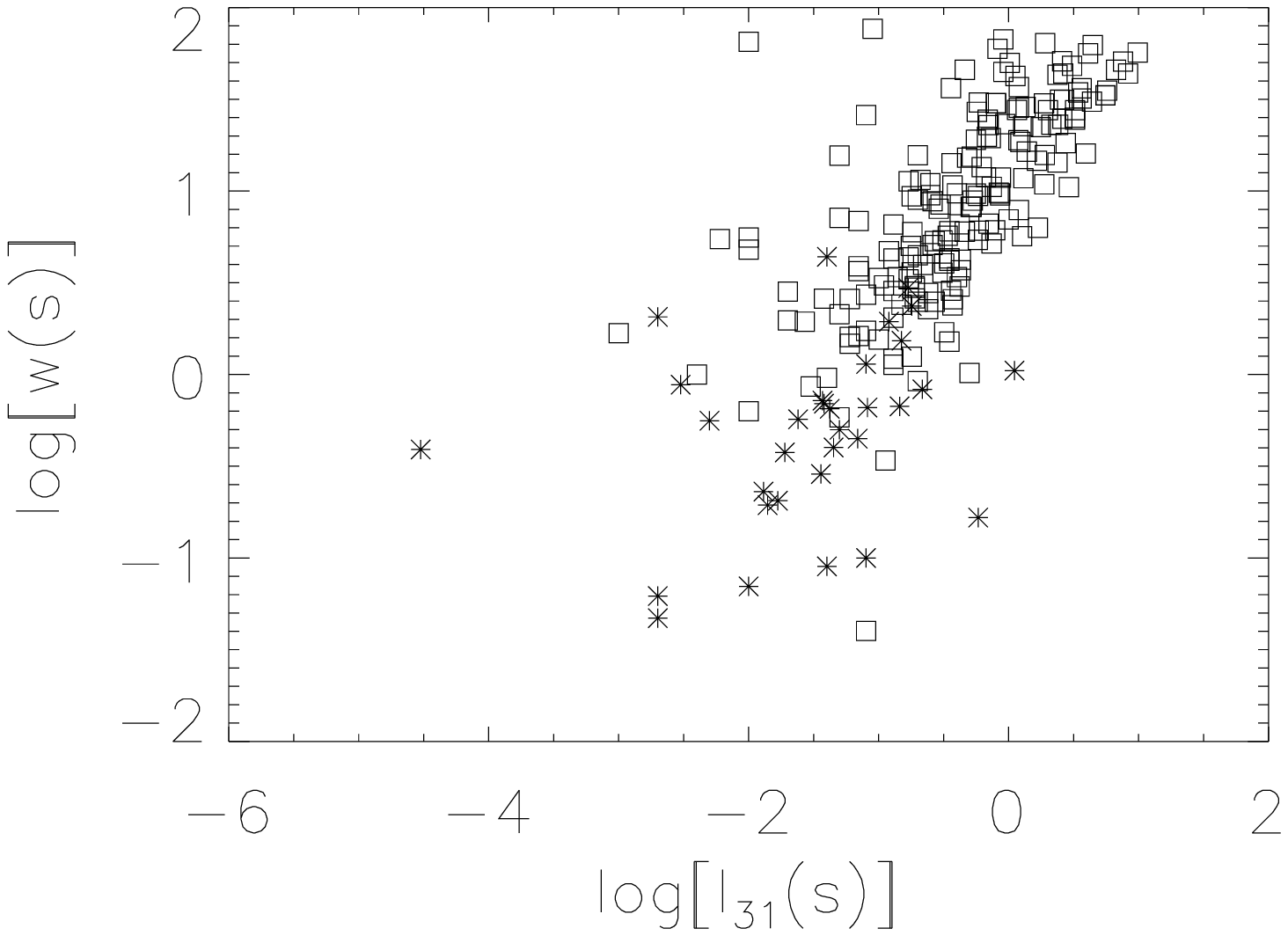}
  \includegraphics[height=.27\textheight]{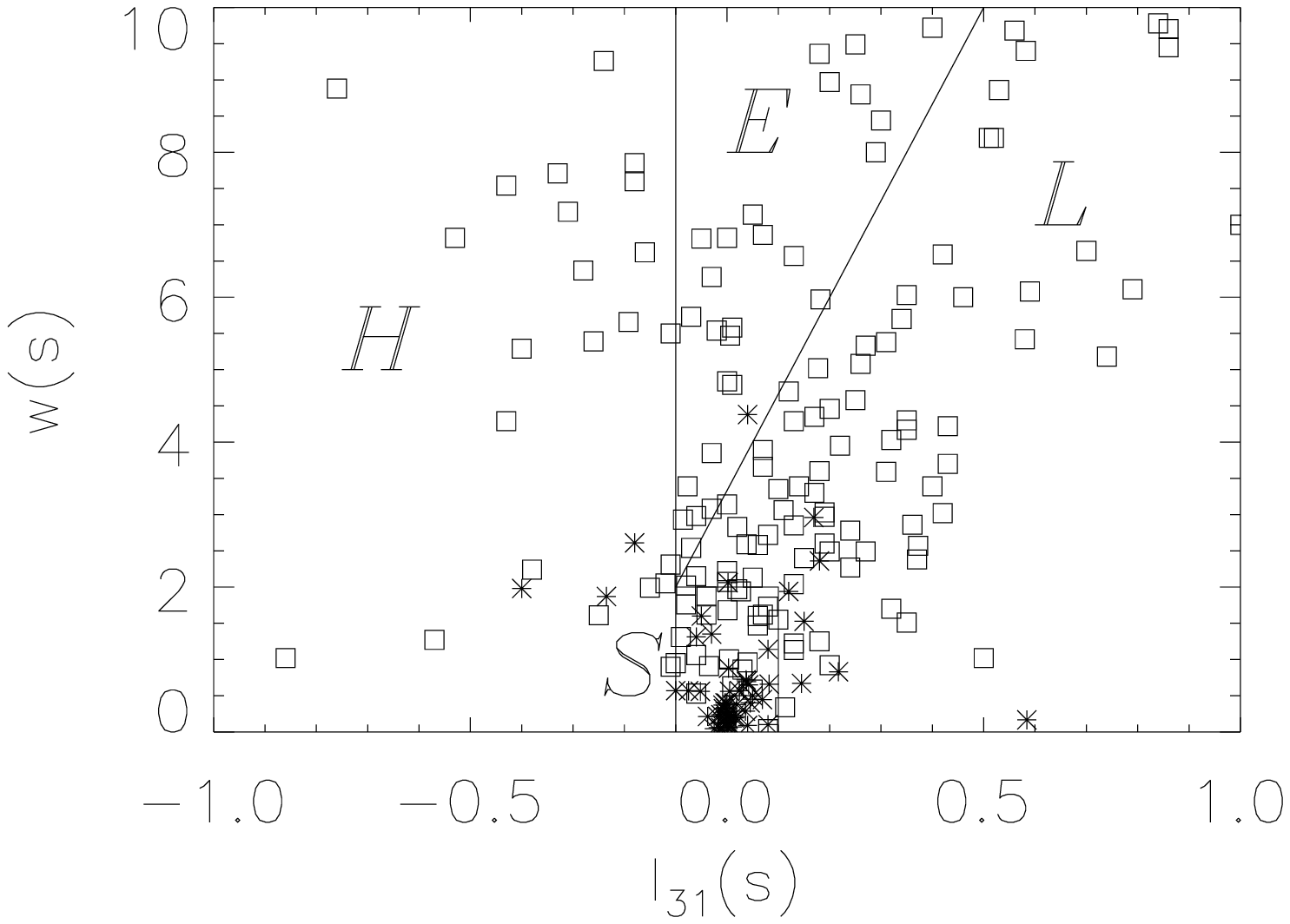}
  \caption{(a) Pulse duration vs. pulse peak lag for Long (square) and Short (*) GRBs in logarithmic units, showing the strong correlation between these parameters. (b) A subset of these data types in linear units, indicating that some pulses have negative lags. Many of these lag measurements are consistent with pulse lags near $l_{31}=0$, but some have abnormally-negative values. Short GRB pulses are typically found in region $S$ (short lags, short durations), while normal Long GRB pulses are found in region $L$ (following the correlation shown in Figure 2(a)). Region $H$ contains Long GRB pulses that have extremely negative lags and which might be marred by hidden pulses, and Region $E$ contains Long GRB pulses that have characteristics potentially associated with external shocks (short pulse lags coupled with long pulse durations).}
\end{figure}

\begin{figure}
  \includegraphics[height=.27\textheight]{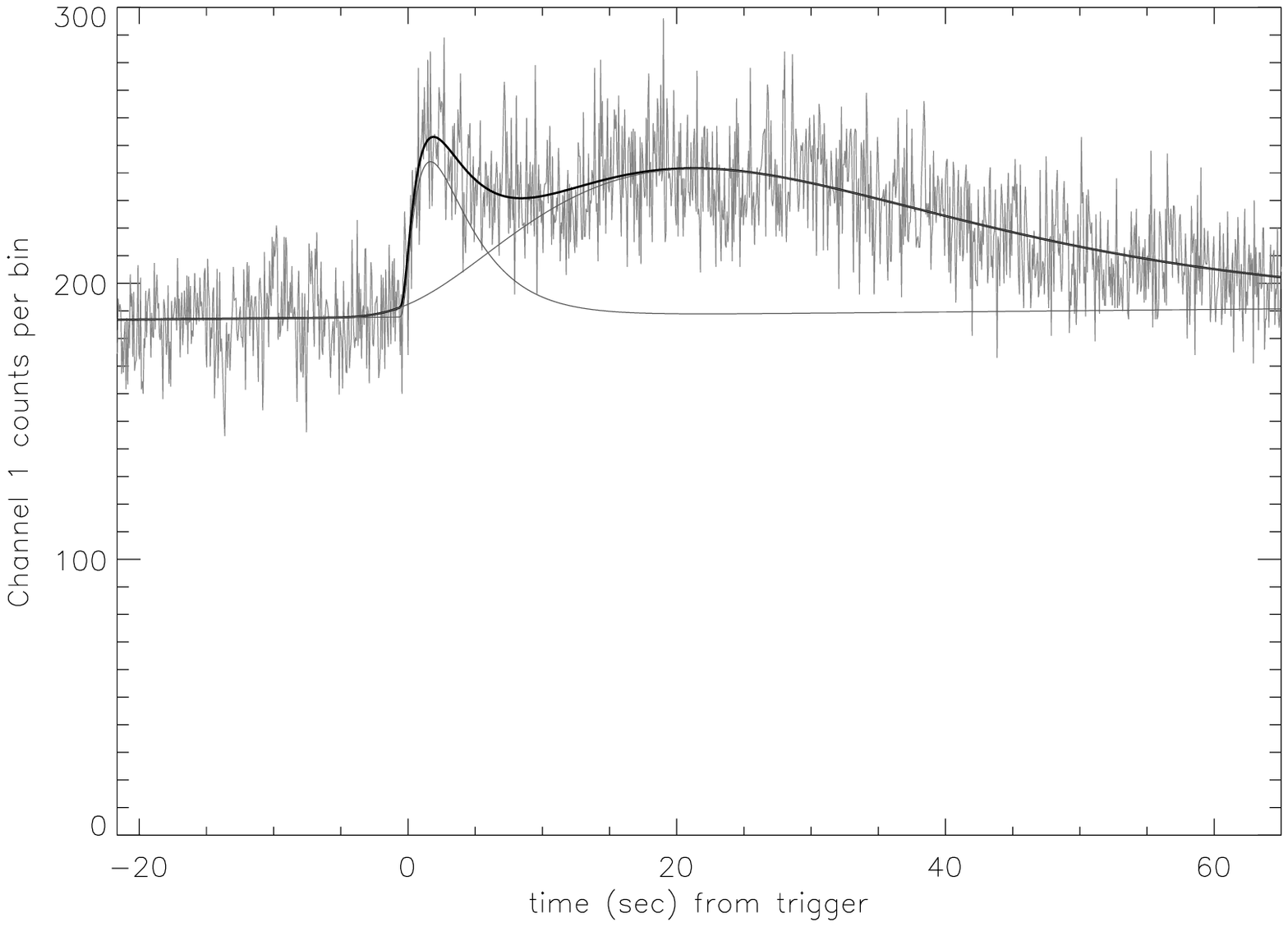}
  \includegraphics[height=.27\textheight]{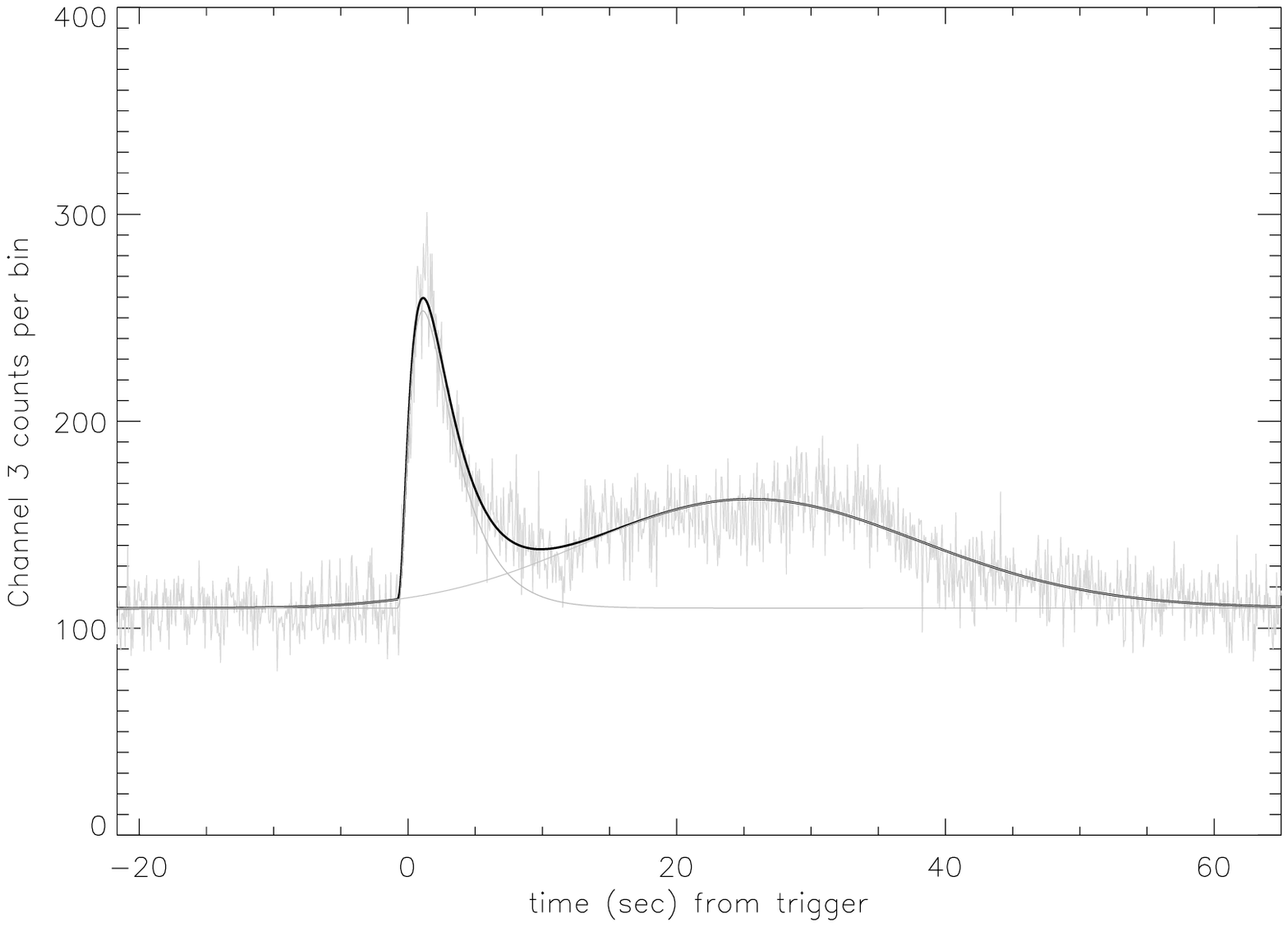}
  \caption{Example of a negative-lag pulse (BATSE Trigger 1807): (a) channel 1, and (b) channel 3. The second fit pulse (with a measured lag of -4.2 s) appears to be composed of two pulses, with the second one being spectrally harder than the first. Merged pulses such as this would explain many of the pulses having negative lags.}
\end{figure}

Figure 4 demonstrates strong anti-correlations between pulse duration and pulse peak flux for both Long and Short GRB pulses. Both Long and Short GRB pulses exhibit anti-correlations, but these relations have different slopes: using a model where $\log p_{256} = A \log w + B$, we find that $A_{\rm Long}=-0.27$ (with correlation coefficient $R=0.36$) and $A_{\rm Short} = -0.67$ ($R=0.64$). Different correlative relations could indicate different pulse mechanisms for the two burst classes. Figure 4 also demonstrates an anti-correlation between spectral hardness and pulse duration for Long GRB pulses. Long, low-luminosity pulses are softer than short, luminous pulses, supporting the argument that intrinsic properties dominate over cosmological effects.

\begin{figure}
  \includegraphics[height=.27\textheight]{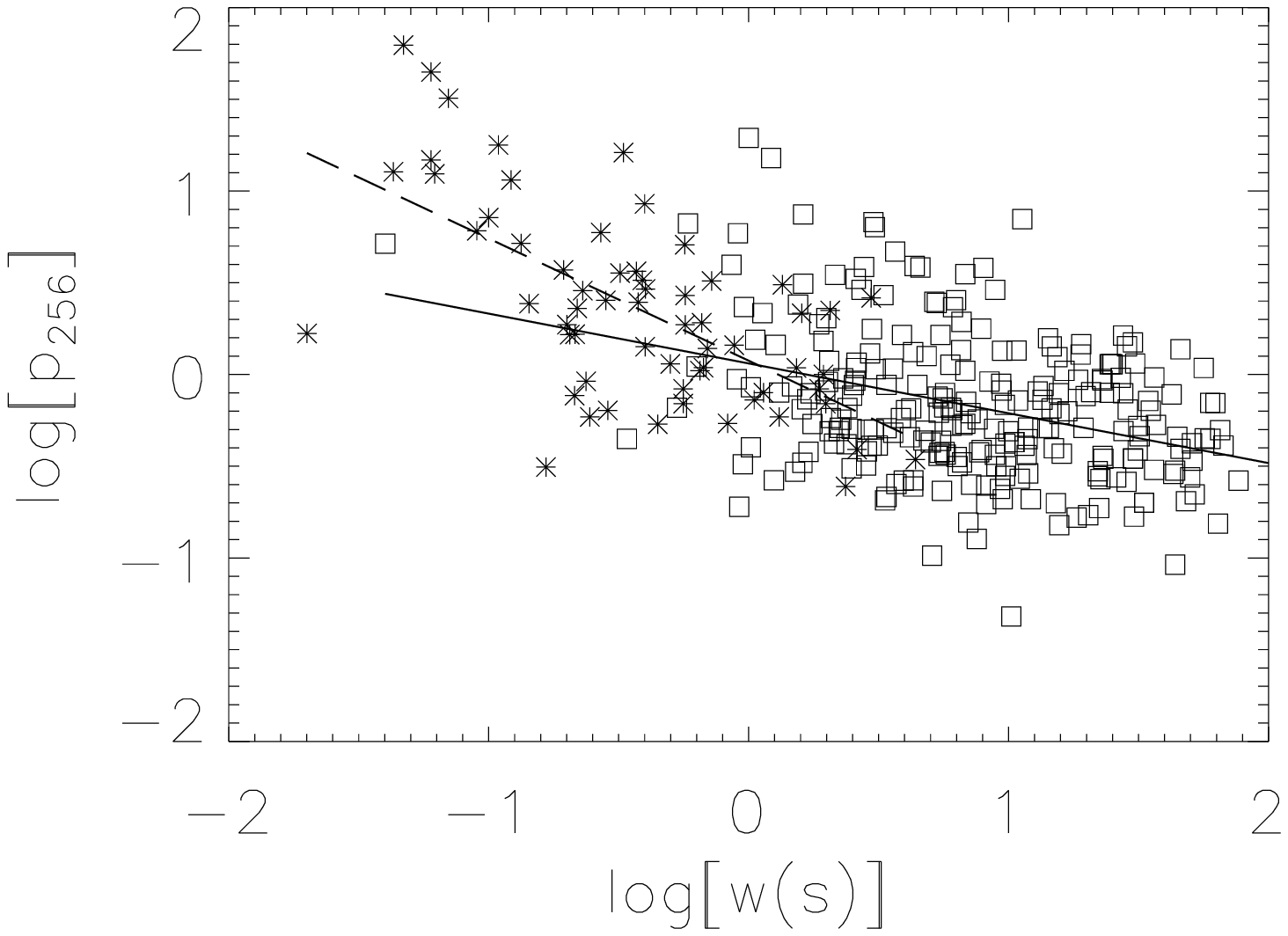}
  \includegraphics[height=.27\textheight]{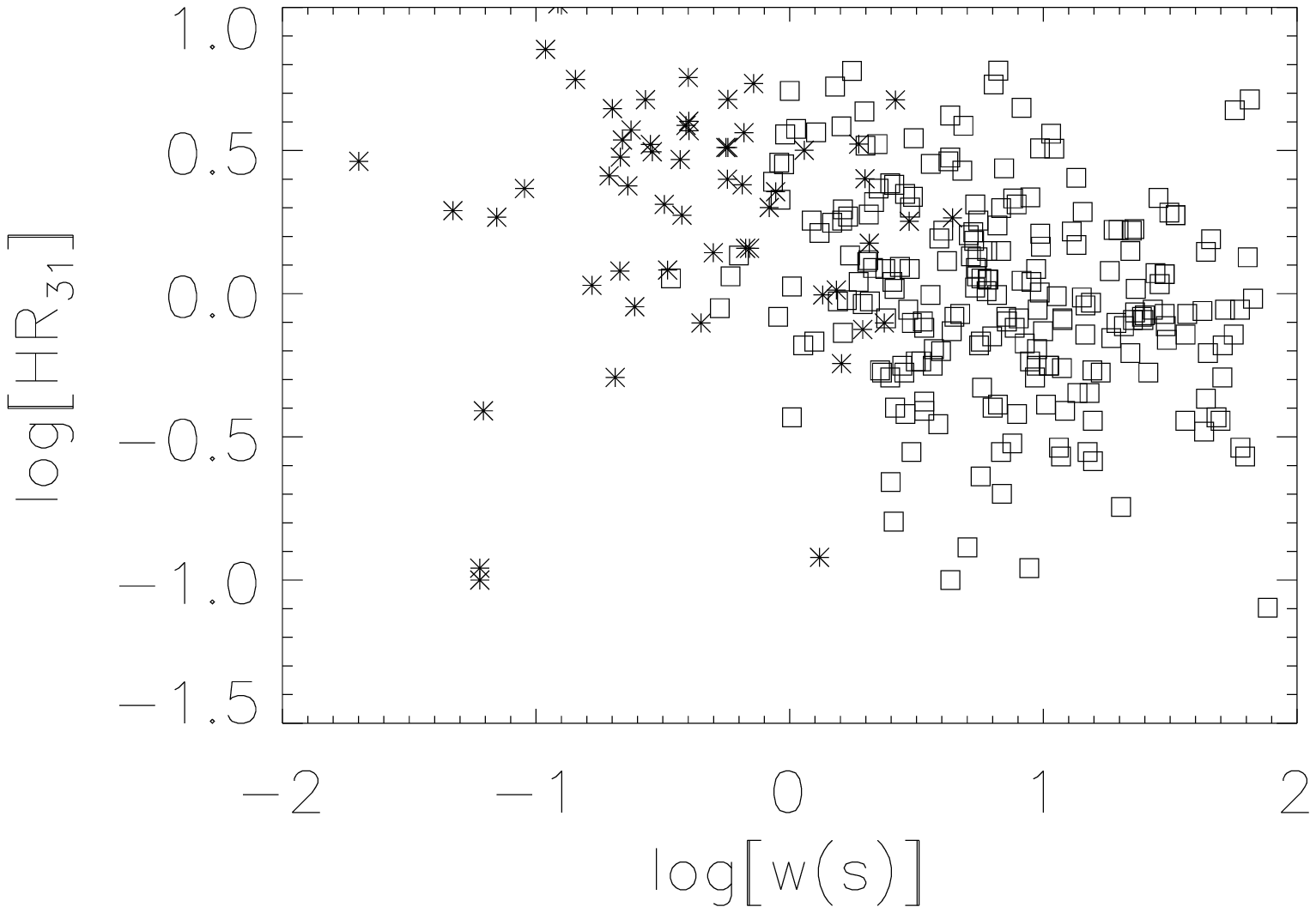}
  \caption{(a) 256 ms peak flux vs.\ pulse duration for Long (square) and Short (*) GRBs. Best-fit lines are given for Long (solid line) and Short (dashed line) relationships. (b) Spectral hardness vs.\ pulse duration for Long (square) and Short (*) GRBs.}
\end{figure}

Figure 5a demonstrates a correlation between Long GRB pulse duration and pulse asymmetry: shorter pulses tend to be more symmetric than longer ones. The asymmetries of Short GRB pulses are not plotted, since the short durations of these pulses (often only a few temporal bins wide) tend to make them be observed as being very symmetric or very asymmetric. Since duration is a luminosity indicator for Long GRB pulses, both spectral hardness and pulse asymmetry are also luminosity indicators.

\begin{figure}
  \includegraphics[height=.27\textheight]{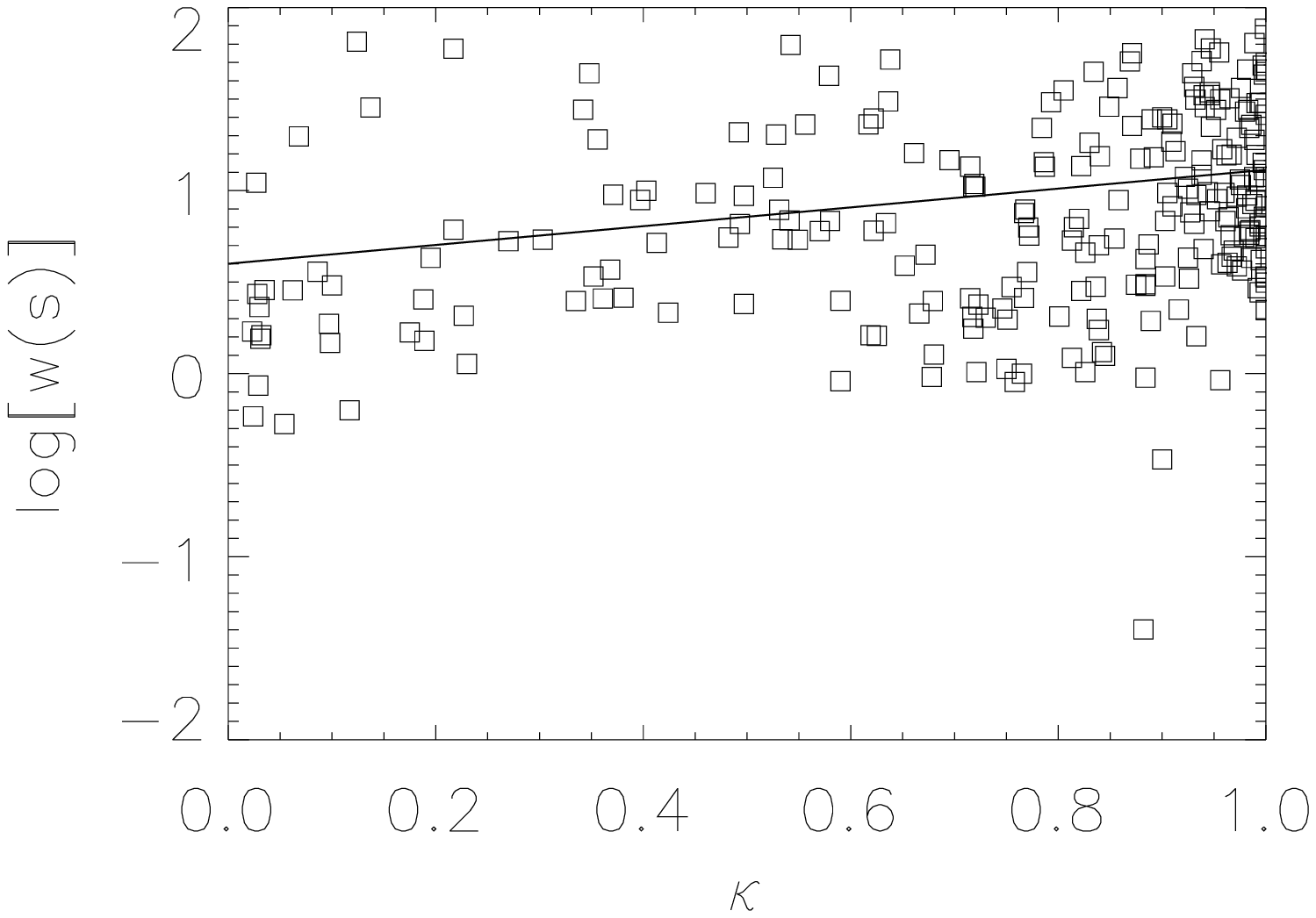}
  \includegraphics[height=.3\textheight]{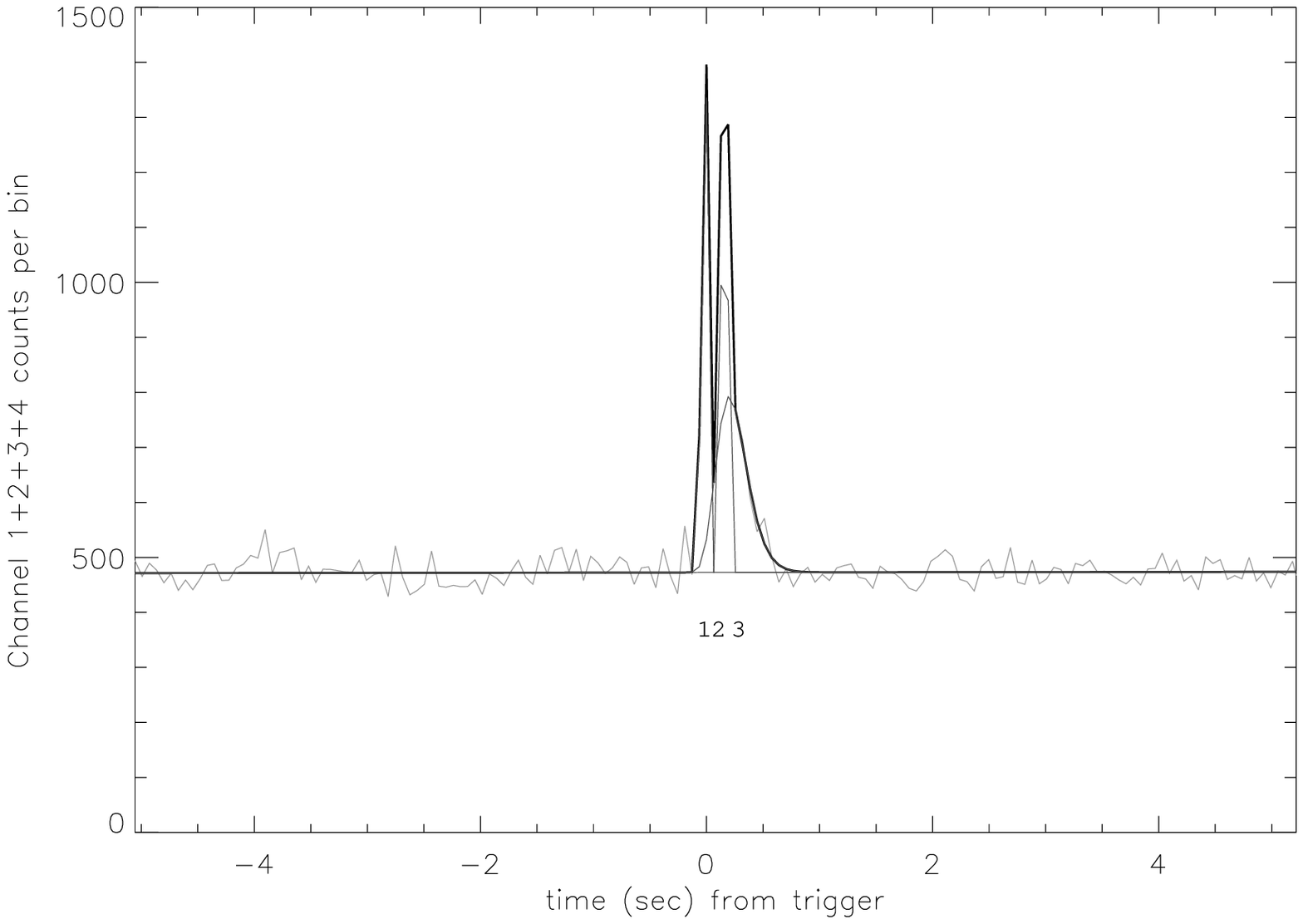}
  \caption{(a) Pulse duration vs.\ asymmetry for Long (square) GRBs, along with a best-fit line. (b) Summed four-channel pulse fits to BATSE trigger 0809, a Short GRB having potential characteristics of the Long GRB class (pulse 1 lag = 0.679 s, pulse 2 lag = 0.368 s, and pulse 3 lag = 0.352 s).}
\end{figure}

The correlative properties of GRB pulses are summarized in Table 1 (Long GRB pulses) and Table 2 (Short GRB pulses). Although Short GRB pulse properties are difficult to accurately measure with 64 ms resolution, there is evidence to suggest that Short GRB pulse properties are not as correlative as Long GRB pulse properties. Such a result could provide a mechanism for more accurate GRB classification. Similarly, pulse property correlations (or a lack thereof) could well provide constraints necessary for constructing more accurate models of GRB prompt emission for Long and Short GRBs.

\begin{table}
\begin{tabular}{lcccc}
\hline
  & \tablehead{1}{c}{b}{$w$}
  & \tablehead{1}{c}{b}{$P_{256}$}
  & \tablehead{1}{c}{b}{$HR_{31}$}
  & \tablehead{1}{c}{b}{$\kappa$}   \\
\hline
$l_{31}$ & ${\bf 2.3\times 10^{-19}}$ & ${\bf ^\dagger 4.8\times 10^{-3}}$ & ${\bf ^\dagger 2.8\times 10^{-3}}$ & $1.1\times 10^{-1}$\\
$w$ & --- & ${\bf ^\dagger 8.8\times 10^{-7}}$ & ${\bf ^\dagger 4.2\times 10^{-7}}$ & ${\bf 1.1\times 10^{-7}}$\\
$P_{256}$ & --- & --- & $1.7\times 10^{-1}$    & ${\bf ^\dagger 5.3\times 10^{-3}}$\\
$HR_{31}$ & --- & --- & ---    & ${\bf ^\dagger 3.3\times 10^{-3}}$\\
\hline
\end{tabular}
\caption{Long GRB pulse correlations. Spearman rank order correlation probabilities that the pulse characteristics in question are random. Anti-correlations are indicated by a $\dagger$, and a significant correlation or anti-correlation is indicated in \bf{boldface}.}
\label{tab:a}
\end{table}

\begin{table}
\begin{tabular}{lcccc}
\hline
  & \tablehead{1}{c}{b}{$w$}
  & \tablehead{1}{c}{b}{$P_{256}$}
  & \tablehead{1}{c}{b}{$HR_{31}$}
  & \tablehead{1}{c}{b}{$\kappa$}   \\
\hline
$l_{31}$ & $2.1\times 10^{-1}$ & $^\dagger 6.6\times 10^{-2}$ & $^\dagger 2.2\times 10^{-2}$ & $4.4\times 10^{-1}$\\
$w$ & --- & ${\bf ^\dagger 7.0\times 10^{-7}}$ & $^\dagger 1.3\times 10^{-1}$ & $^\dagger 8.2\times 10^{-2}$\\
$P_{256}$ & --- & --- & $7.3\times 10^{-2}$    & $7.9\times 10^{-1}$\\
$HR_{31}$ & --- & --- & ---    & $2.3\times 10^{-2}$\\
\hline
\end{tabular}
\caption{Short GRB pulse correlations.}
\label{tab:b}
\end{table}

Despite the insights gained from these pulse relationships, there is still ambiguity in the definition of Short and Long GRBs as obtained from bulk emission and pulse properties. Take, for example, BATSE trigger 0809 (Figure 5b): this burst is hard with a $T_{90}$ duration of less than two seconds, and would normally be classified as a Short GRB. However, the pulse properties of pulse duration, pulse peak flux, and pulse peak lag correlate with one another in a way that is consistent with it being a Long GRB. Thus, where many observers are now making claims to expand the Short GRB definition to reclassify many Long bursts as Short, we have found an ambiguous Short GRB that might actually be a Long GRB solely because it is lacking inter-pulse durations.

\section{Conclusions}

A large GRB pulse sample verifies that each GRB is characterized by its own lag. The lag of a Long GRB pulse is directly related to its duration, and both of these properties anti-correlate with the pulse peak luminosity. Via correlations with pulse peak lag and pulse peak duration, it also appears that pulse spectral hardness and pulse asymmetry are luminosity indicators. These relationships hold for most Long GRB pulses, but perhaps exclude long pulses with short, intense pulse peaks (external shocks?).

Short GRB pulses do not reflect the same correlative behaviors as those found for Long burst pulses, although this result might partly reflect the limited temporal resolution used in this study. Perhaps the most interesting behavior of Short burst pulses is an anti-correlation between pulse duration and pulse peak flux which might indicate a relationship between pulse duration and pulse peak luminosity.

There is a need for better measuring and understanding GRB pulse properties. In principle, the bulk characteristics of the prompt emission can be derived from knowledge of pulse properties and the pulse decomposition of a burst. The converse, however, is not true: we cannot infer the basic characteristics of the pulses, nor can we understand the relationship between bulk and (more fundamental) pulse properties, without explicit study of the constituent pulses.


\begin{theacknowledgments}
We thank Tom Loredo, Rob Preece, Tim Giblin, Chris Fragile, and Jay Norris for valuable discussions. R. Cumbee acknowledges financial support from the South Carolina Space Grant program and faculty support from Dr. J. Myers of Francis Marion University. We also gratefully thank USRA and the conference poster competition judges.
\end{theacknowledgments}



\bibliographystyle{aipprocl} 

\bibliography{pulse_props}

\IfFileExists{\jobname.bbl}{}
 {\typeout{}
  \typeout{******************************************}
  \typeout{** Please run "bibtex \jobname" to optain}
  \typeout{** the bibliography and then re-run LaTeX}
  \typeout{** twice to fix the references!}
  \typeout{******************************************}
  \typeout{}
 }

\end{document}